\documentstyle[11pt,aaspp4,psfig]{article}

\begin{document}

\title{Hydrodynamic Effects on the QPO Frequencies of Accreting
Compact Objects}

\author{Dimitrios Psaltis}
\affil{\footnotesize Harvard-Smithsonian Center for Astrophysics, 60 Garden
St., Cambridge, MA 02138\\
and \\
Center for Space Research,
Massachusetts Institute of Technology, Cambridge, MA 02139;\\
demetris@space.mit.edu}

\begin{abstract}
The variability properties of accreting compact objects predicted by
dynamical models are characterized by a number of distinct frequencies
that are specific to each model.  Because the accretion disks around
neutron stars and black holes are hydrodynamic flows, these modulation
frequencies cannot be strictly equal to test particle frequencies. I
discuss several implications of hydrodynamic corrections to the
modulation frequencies predicted by dynamical models. Finally, I show
that the recent detection of yet a third kHz QPO in three neutron-star
systems favors a previously developed model that attributes the
various QPO frequencies to fundamental general relativistic
frequencies in the accretion flow.
\end{abstract}


\centerline{Submitted to {\em The Astrophysical Journal Letters.}}

\section{INTRODUCTION}

The accretion flows around compact objects are turbulent and are,
therefore, expected to be highly variable. The frequencies that
correspond to the dynamical timescales can be as high as $\simeq
1$~kHz and depend strongly on the distance from the central object.
For example, for a Kerr spacetime and in the limit of slow rotation,
the Keplerian orbital frequency at a radius $r$ scales as $\sim
r^{-3/2}$, the periastron precession frequency as $\sim r^{-5/2}$,
whereas the nodal precession frequency (if it is caused by frame
dragging) scales as $\sim r^{-3}$ (see, e.g., Perez et al.\ 1997;
Stella, Vietri, \& Morsink 1999). Because of such strong dependences,
it is surprising that the observed power-density spectra of accreting
systems are characterized by a distinct number of frequencies in the
form of relatively coherent ($\Delta \nu/\nu\gtrsim 10^{-2}$)
quasi-periodic oscillation (QPO) peaks.

Modeling theoretically the presence of multiple QPO peaks is
challenging. Current models seek to identify characteristic radii in
the accretion flow that pick only a small range of frequencies and
hence produce narrow QPO peaks in the power spectra. Such radii
include the radius of the innermost stable circular orbit (which is
important in the diskoseismic models of, e.g., Nowak \& Wagoner 1991;
Perez et al.\ 1997; Silbergleit \& Wagoner 2000), the magnetospheric
radius (Alpar \& Shaham 1985, Strohmayer et al.\ 1996), or the
sonic-point radius (Miller, Lamb, \& Psaltis 1998).  The simultaneous
presence of low- and high-frequency QPOs is then attributed either to
multiple characteristic radii in the accretion flow (as discussed,
e.g., in Miller et al.\ 1998; Psaltis et al.\ 1999b) or to different
characteristic frequencies occurring at the same region in the flow,
as in the case of the diskoseismic models (see Wagoner 1999) or the
relativistic models of Stella et al.\ (1999) and Psaltis \& Norman
(2000).  In the first class of models, the observed tight correlations
between the different QPO frequencies (see, e.g., Psaltis et al.\
1999b; Psaltis, Belloni, \& van der Klis 1999a) may be accounted for
by combining the dependence of each QPO frequency on the mass
accretion rate (see, e.g., Psaltis et al.\ 1999b). In the latter case,
frequency correlations can be easily reproduced since all QPO
frequencies are determined almost entirely by a single parameter
(e.g., the characteristic transition radius; Psaltis \& Norman 2000).

A characteristic signature of each QPO model is the spectrum of QPO
frequencies that it predicts beyond the ones currently detected (see,
e.g., Miller 2000). For this reason, the inadequacy of current data to
distinguish between different models can be overcome with the
detection of additional QPO peaks. Even though some attempts for
detecting expected QPO peaks at different frequencies have given
negative results (M\'endez \& van der Klis 2000), a third kHz QPO has
been recently discovered in three atoll sources (Jonker, M\"endez, \&
van der Klis 2000). This new QPO is significantly weaker than the
other kHz QPOs and has a frequency that is $\simeq 60$~Hz higher than
that of the previously known lower kHz QPO.

Motivated by this detection, I discuss in this {\em Letter} the
importance of the hydrodynamic corrections to the predicted QPO
frequencies that are introduced by the previously developed dynamical
model that identifies the various QPO frequencies with general
relativistic frequencies in the accretion disk (Psaltis \& Norman
2000). I then use this property to show that the model can account for
the observed properties of the newly discovered QPO.

\section{HYDRODYNAMIC EFFECTS ON THE PREDICTED QPO FREQUENCIES}

The high frequencies that characterize the observed QPOs in neutron
star sources imply that the accretion flows around them must be
modulated at the dynamical timescale. A simple estimate of the three
characteristic dynamical timescales at distance $r$ away from a
compact object is given by the three test-particle frequencies, i.e.,
the orbital frequency in the azimuthal direction, the epicyclic
frequency in the radial direction, and the oscillation frequency in
the vertical direction. For a slowly rotating compact object these
frequencies are (see, e.g., Perez et al.\ 1997)
\begin{equation}
   \Omega^2=\frac{M}{r^3[1+\alpha_*(M/r)^{3/2}]}\;,
\label{eq:omegak}
\end{equation}
\begin{equation}
   \kappa^2=\Omega^2(1-6M/r)\;,
\label{eq:kappa}
\end{equation}
and
\begin{equation}
   \Omega_\perp^2=\Omega^2[1-4\alpha_*(M/r)^{3/2}]\;,
\label{eq:omegap}
\end{equation}
respectively, where $M$ is the mass and $\alpha_*$ the specific
angular momentum per unit mass of the compact object, and I have set
the fundamental constants to $c=G=1$. The resulting orbital frequency
($\Omega/2\pi$), periastron precession frequency
[$(\Omega-\kappa)/2\pi$], and nodal precession frequency
[$(\Omega-\Omega_\perp)/2\pi$] were shown to be comparable to and
follow similar correlations with the three observed QPO frequencies
(Stella \& Vietri 1998, 1999; Stella et al.\ 1999).

An accretion disk, however, is a hydrodynamic flow and hence the
test-particle frequencies~(\ref{eq:omegak})--(\ref{eq:omegap}) cannot
be strictly equal to the frequencies at which the flow is modulated.
For example, the frequencies of the trapped $g-$ and $c-$modes in the
model discussed in Wagoner (1999) are determined mostly by the maximum
of the radial epicyclic frequency (eq.~[\ref{eq:kappa}]) and the
nodal-precession frequency ($\Omega-\Omega_\perp$) but also depend
weakly on the local sound speed. In the case of the model discussed in
Psaltis \& Norman (2000), the inner accretion disk is modulated mainly
at three characteristic frequencies,
\begin{equation}
2\pi f_1=\Omega-\Omega_\perp\left[1+\frac{\Omega_{\rm c}^2
   (\Omega_\perp^2-\kappa^2)}{2\Omega_d^2\Omega_\perp^2+
   2(\kappa^2-\Omega_\perp^2)^2}\right]\;,
\end{equation}
\begin{equation}
2\pi f_2=\Omega-\kappa\left(1+\frac{\Omega_c^2}{2\kappa^2}\right)\;,
\label{eq:f2}
\end{equation}
and
\begin{equation}
2\pi f_3=\Omega\;,
\end{equation}
where $\Omega_c\equiv c_{\rm s}/r$, $c_{\rm s}$ is the speed of sound,
and $\Omega_d\sim u_r/r$ is the inverse radial drift timescale at the
transition radius $r$. In the notation of Psaltis \& Norman (2000),
the mode $(m=1,n=0)$ provides the biggest contribution to the peak at
frequencies $f_2$ and $f_3$, whereas the mode $(m=1,n=1)$ provides the
biggest contribution to the peak at frequency $f_1$.

It is important to note here that the corrections to the three
dynamical frequencies introduced by the model are specific to the fact
that they describe the modulation frequencies of the {\em density\/}
in the inner disk and depend on the details of the calculation.  For
example, because of the approximations used, the effects of radial
pressure forces, which will affect the hydrodynamic corrections, have
not been taken explicitly into account. The corrections introduced to
the modulation of the X-ray luminosity will be different in detail but
similar in magnitude, if such effects are taken into account.
However, hydrodynamic corrections are generic to a dynamical model and
have three important implications.

\begin{figure}[t]
\centerline{
\psfig{file=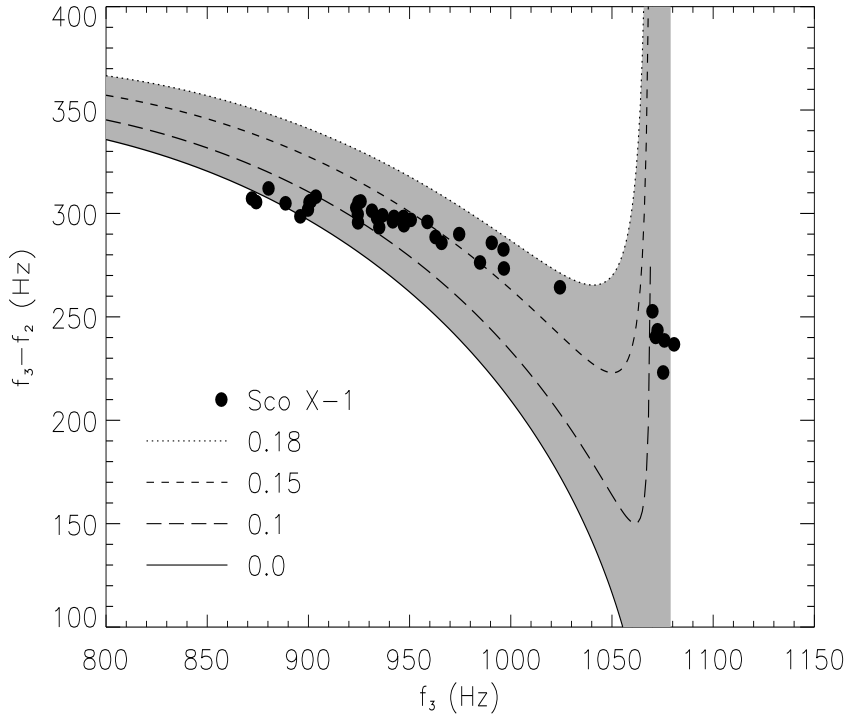,angle=90,height=10.5truecm,width=12truecm}}
\figcaption[]{
The effect of hydrodynamic corrections on the frequency difference
$f_3-f_2$, calculated for a non-rotating compact object of mass $2.05
M_\odot$, a viscosity parameter $\alpha=0.2$, a fractional width of
the transition region of $\delta r/r=0.01$, and different values of
the disk scale-height $h/r$ (see Psaltis \& Norman 2000 for
definitions). The data points are for the neutron-star source Sco~X-1
and correspond to the peak separation of the kHz QPOs and the
frequency of the upper QPO peak (after van der Klis et al.\ 1997).}
\end{figure}

First, the observed QPO frequencies will not be strictly equal to the
general relativistic frequencies; the corrections can be as large as
$10-30$\%, depending on the local sound speed and the viscosity in the
accretion flow (see, e.g., Psaltis \& Norman 2000). This property
resolves entirely the small but statistically significant discrepancy
in the fits of the uncorrected general relativistic frequencies to
observations (Psaltis et al.\ 1999b; Markovi\'c \& Lamb 2000). An
example is shown in Figure~1, where the predicted frequency difference
$f_3-f_2$ is compared to the changing peak separation of the kHz QPOs
in the neutron-star source Sco~X-1 (van der Klis et al.\ 1997), for
standard values of the model parameters. The observed decrease of the
kHz QPO peak separation with increasing frequency is slower than
predicted by the model but can be accounted for with a simultaneous
small increase of the disk scale height.  The resulting frequencies
are in good agreement with the data without the need for introducing
characteristic frequencies of elliptical orbits (as originally
suggested by Stella \& Vietri 1998), which are unlikely to be
sustained in an accretion disk. Note here that the results plotted in
Figure~1 are for a non-rotating compact object and were not obtained
from formally fitting the data, but are shown to illustrate the effect
of hydrodynamic corrections. However, detailed comparisons with data
of the dynamical frequencies calculated for realistic neutron-star
spacetimes were shown to affect only slightly the conclusion
(Markovi\'c \& Lamb 2000).

Second, the centroid frequencies of QPO peaks produced by resonances
that correspond to harmonics of general relativistic frequencies are
not exactly harmonically related. This is true, because the
hydrodynamic corrections are different for the different modes and
hence the ratio of the frequencies of the modes is not an integer
number but depends on the properties of the flow. Note, however, that
if the presence of additional harmonics besides the fundamental is
caused by the non-sinusoidal profile of the modulation of the X-ray
flux and not by the presence of higher-order modes, then the frequency
ratio will be an integer. This might lead to the presence of multiple
peaks at nearby frequencies with possibly different FWHMs, as for
example observed in the case of Cyg~X-1 by Nowak (2000).

Finally, different modes may contribute power to the same QPO peak but
at slightly different centroid frequencies, again because of the
different hydrodynamic corrections. For example, all the ($m=1,n\ge
0$) modes in the analysis of Psaltis \& Norman (2000) contribute to
the variability power at frequencies comparable to $\sim f_2$, i.e.,
to the lower kHz QPO (the contribution of the modes with $n>0$ is
small and was not explicitly pointed out in the original analysis,
even though it was present). Indeed, expanding the general expression
for the response function (eq.~[27] in Psaltis \& Norman 2000) for the
($m=1,n=1$) mode and at frequencies $2\pi f\sim \Omega-\kappa$, it
simplifies to
\begin{equation}
   A_{11}\simeq\left\{\left[1+\frac{\Omega_{\rm c}^2}{4\omega_{\rm b}}
   \frac{\Omega_{\rm d}}{(2\pi f-\Omega+\kappa)^2}\right]^2
   +\left[-\frac{\Omega_\perp^2-\kappa^2}{\kappa\omega_{\rm b}}+
   \left(\frac{\Omega_{\rm c}}{2\omega_{\rm b}}\right)
   \frac{\Omega_{\rm c}}{2\pi f-(\Omega-\kappa)}\right]^2\right\}^{-1/2}\;,
\label{eq:A11}
\end{equation}
when $(\delta r/r)(\kappa/\Omega_{\rm c})^2\ll
1$. Equation~(\ref{eq:A11}) describes a narrow resonant mode centered
at a frequency
\begin{equation}
   2\pi f_{2b}\simeq \Omega-\kappa\left[1-\frac{\Omega_{\rm c}^2}
     {2(\Omega_\perp^2-\kappa^2)}\right]\;,
\label{eq:f2b}
\end{equation}
and with a FWHM of
\begin{equation}
\delta f_{2b}\simeq\frac{\sqrt{3}}{2\pi}\omega_{\rm b}
   \left(\frac{\Omega_{\rm c}^2}
     {\Omega_\perp^2-\kappa^2}\right)\;,
\end{equation}
where $\omega_{\rm b}=(r/\delta r)\Omega_{\rm d}$ and $\delta r/r$ is
the fractional width of the transition region. This resonance is to
be compared to the corresponding resonance of the ($m=1,n=0$) mode,
which is described by (see, eq.~[35] of Psaltis \& Norman 2000)
\begin{equation}
A_{10}\simeq\left\{1+\left[\frac{\kappa}{\omega_{\rm b}}+
   \left(\frac{\Omega_{\rm c}}{2\omega_{\rm b}}\right)
   \frac{\Omega_{\rm c}}{2\pi f-(\Omega-\kappa)}\right]^2\right\}^{-1/2}\;,
   \label{eq:A10}
\end{equation}
has a peak at frequency $f_2$ and a FWHM of
\begin{equation}
\delta f_{2}\simeq \frac{\sqrt{3}}{2\pi}\omega_{\rm b}
   \left(\frac{\Omega_{\rm c}^2}{\kappa^2}\right)\;,
\end{equation}
when $(\delta r/r)(\kappa/\Omega_{rm c})^2\ll 1$.

\begin{figure}[t]
\centerline{
\psfig{file=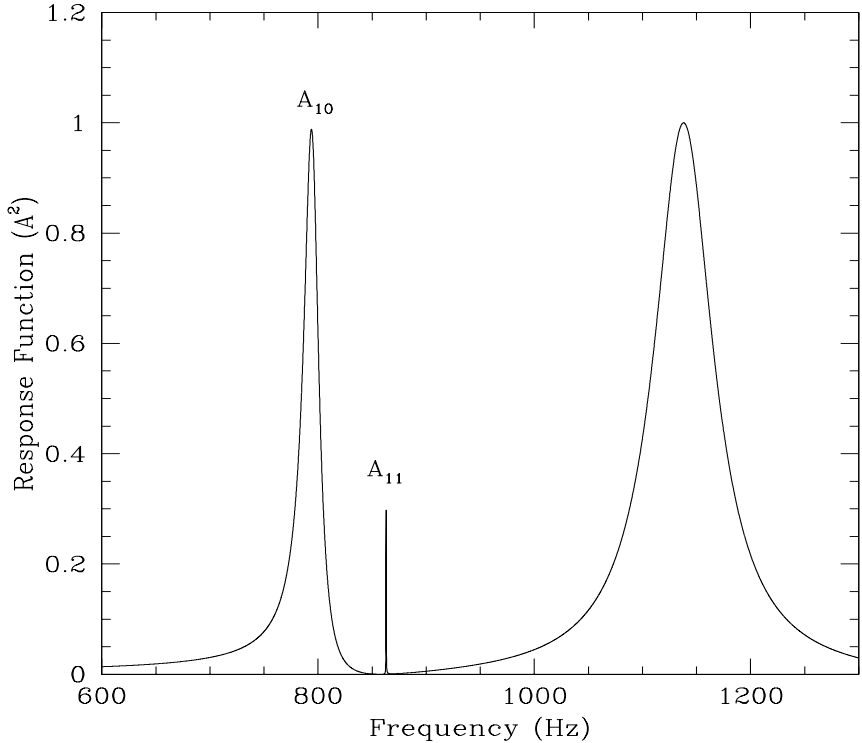,angle=0,height=10.5truecm,width=12truecm}}
\figcaption[]{
The response function for modes $(m=1,n=0)$ and ($m=1,n=1$) at high
frequencies, calculated for $\Omega/2\pi=1138$~kHz,
$\kappa/2\pi=280$~Hz, $(\Omega-\Omega_\perp)/2\pi=50$~Hz, $\Omega_{\rm
c}/2\pi=200$~Hz, and $\omega_{\rm b}/2\pi=50$~Hz. The parameters
correspond to typical values of the hydrodynamic corrections in an
accretion disk (see eq.~[14]) and were chosen to reproduce the
observed power spectrum of the source 4U~1728$-$34 (Jonker et al.\
2000).}
\end{figure}

Figure~2 shows the resulting response of these lowest-order modes at
high frequencies. As it is also apparent from equations~(\ref{eq:f2})
and (\ref{eq:f2b}), the two peaks at frequencies comparable to $\simeq
f_2$ are displaced relative to each other and, in the particular model
considered here, their separation is equal to
\begin{equation}
f_{2b}-f_2\simeq \frac{\Omega_c^2}{4\pi\kappa}
   \left(1+\frac{\kappa^2}{\Omega_\perp^2-\kappa^2}\right)\;.
\label{dif}
\end{equation}
As a result, two QPOs at frequencies comparable to that of the lower
kHz QPO may appear in the power spectra of accreting compact objects.
Note here that the two frequencies are identical in the limit
$\Omega_c\rightarrow 0$, i.e., for a purely kinematic model.

The amplitudes of the two QPOs will depend on the power-spectrum of
the driving perturbations, which might be different in the two modes
that produce the two QPOs, even though their frequencies are very
similar. (The maximum of the response function does {\em not\/}
provide an estimate of the amplitudes of the QPOs.)  However, even for
a comparable amplitude of perturbations in both modes, the resonance
described by $A_{10}$ will produce the dominant QPO peak (see also
Psaltis \& Norman 2000), whereas the resonance described by $A_{11}$
will produce a weaker sideband, as Figure~2 suggests.  In the absence
of any additional broadening mechanisms, the FWHM of the two lower kHz
QPOs (at frequencies $f_2$ and $f_{\rm 2b}$) will be significantly
smaller than the FWHM of the upper kHz QPO (at frequency
$f_3$). However, the two lower kHz QPOs may have a difference of up
to $\sim (\Omega_\perp/\kappa)^2\lesssim 15$ in their relative FWHM.

\section{DISCUSSION}

In the previous section, I discussed a number of implications of the
hydrodynamic corrections introduced to the predicted QPO frequencies
by dynamical models. Moreover, I showed that the previously developed
model that attributes the observed QPOs to general relativistic
frequencies in the accretion flow can account for the presence of
multiple kHz QPOs, as recently observed by Jonker et al.\ (2000).  In
this section I will concentrate to the case of the well studied source
4U~1728$-$34 (Jonker et al.\ 2000) and discuss the implications of the
newly observed QPOs for various models. In this source, four
QPOs are observed simultaneously at the following frequencies:
$\nu_1=41.5\pm 0.2$~Hz (the low-frequency QPO), $\nu_2=795$~Hz (the
lower kHz QPO), $\nu_{2b}=859\pm 2$~Hz (the newly discovered QPO), and
$\nu_3=1138\pm 2$~Hz (the upper kHz QPO).

According to beat-frequency models (see, e.g., Strohmayer et al.\
1996; Miller et al.\ 1998), the highest QPO frequency $\nu_3$ is that
of the Keplerian frequency of a characteristic radius in the accretion
disk and the difference $\nu_3-\nu_2$ is comparable to the spin
frequency of the neutron star. In fact, oscillations during Type~I
X-ray bursts have been detected from 4U~1728$-$34 at a frequency of
$\simeq 363$~Hz, which are also thought to correspond to the spin
frequency of the neutron star (Strohmayer et al.\ 1996). In this
picture, the presence of the newly discovered QPO at a frequency that
is $\simeq 64$~Hz higher than $\nu_2$ might be accounted for if the
radiation that emerges from the neutron star is modulated not only at
the stellar spin frequency, but also at a frequency equal to $\simeq
64$~Hz, introducing sidebands to the stellar spin (see, however, Alpar
1986 for an alternative way of generating sidebands in a
beat-frequency model, because of the radial drift of the accreting
material). As pointed out by Jonker et al.\ (2000), this latter
frequency is not equal to the frequency of the observed QPO at $\nu_1$
and therefore the accretion flow needs to produce modulations at {\em
two\/} distinct, unrelated, low frequencies. One such low frequency is
the nodal precession frequency $(\Omega-\Omega_\perp)/2\pi$.  Note,
however, that, as also discussed by Psaltis et al.\ (1999b) for other
similar cases, the QPO at frequency $\nu_1$ cannot be caused by
frame-dragging effects if the spin-frequency of the neutron star is
equal to $\sim 363$~Hz. (In the relativistic model described in \S2
the frequency $\nu_1$ is related to the frame-dragging frequency, but
the peak separation of the kHz QPOs is not equal to the spin frequency
of the neutron star.)  This can also be extended to the case of the
$\simeq 64$~Hz frequency, which is even higher than $\nu_1$. Indeed,
such a high nodal precession frequency would require an
unrealistically high neutron-star moment of inertia of (see, e.g.,
Psaltis et al.\ 1999b)
\begin{equation}
\frac{I_{45}}{M/M_\odot}\gtrsim
   3.1\left(\frac{\nu}{64~\mbox{Hz}}\right) \left(\frac{\nu_{\rm
   spin}}{363~\mbox{Hz}}\right)^{-1}
   \left(\frac{\nu_3}{1138~\mbox{Hz}}\right)^{-2}\;,
\end{equation}
where $I_{45}\equiv I/10^{45} g\,cm^3$ is the moment of inertia,
$M/M_\odot$ the mass in solar units, and $\nu_{\rm spin}$ the spin
frequency of the compact object. Even if a frequency of $\simeq 64$~Hz
corresponded to the first overtone of the nodal precession frequency,
the required moment of inertia [$I_{45}/(M/M_\odot)\gtrsim 1.55$]
would be higher than allowed by general relativity for a slowly
rotating star, if any of the current equations of state for
neutron-star matter were valid up to $\simeq 1.5$ times the nuclear
saturation density. As a result, none of the two required low
frequencies can be equal to the frame-dragging frequency, if the spin
frequency of the neutron star in 4U~1728$-$34 is equal to the burst
oscillation frequency.

In the case of the dynamical model described in Psaltis \& Norman
(2000), the presence of multiple QPO peaks in the power-density
spectra is a direct consequence of the fact that in strong-field
general relativity the three dynamical frequencies
(eq.~[\ref{eq:omegak}]--[\ref{eq:omegap}]) are significantly different
from each other. In particular, all the QPO peaks observed so far can
be accounted for by only the two lowest-order modes ($m=1,n=0$) and
($m=1,n=1$).

In this interpretation, the difference $f_{2b}-f_2$ depends mostly on
the radial epicyclic frequency and the inverse sound-crossing time
(see, eq.~[\ref{dif}]). For a standard $\alpha$-disk, the latter is
equal to $\Omega_{\rm c}\simeq (h/r) \Omega_\perp$, where $(h/r)$ is
the scale-height of the disk (see, e.g., Psaltis \& Norman 2000). As a
result, the predicted frequency separation between the two lower-kHz
QPO peaks becomes
\begin{equation}
f_{2b}-f_2 \simeq 66\left(\frac{h/r}{0.12}\right)^2
    \left(\frac{\Omega/2\pi}{1138~\mbox{Hz}}\right)^2
    \left(\frac{\kappa/2\pi}{280~\mbox{Hz}}\right)^{-1}~\mbox{Hz}
\end{equation}
and hence the observed frequencies can be easily accounted for, for
typical values of the model parameters (see also Fig.~2).

Higher-order modes will also contribute to the variability power at
frequencies comparable to $f_2$ and may produce additional structure
to the lower kHz QPO peak. The overall spectrum of observed modes will
depend mostly on the amplitude of the perturbations that drive the
variability outside the characteristic transition radius.  A detailed
analysis of the various predicted resonant frequencies and their
amplitudes is beyond the scope of this paper and will be presented
elsewhere.

\acknowledgments

I am grateful to Peter Jonker and Mariano M\'endez for sharing their
results in advance of publication. I thank Ali Alpar, Deepto
Chakrabarty, Mike Nowak, Feryal \"Ozel, and Michiel van der Klis for
useful discussions and comments on the manuscript.  I also acknowledge
the support of a postdoctoral fellowship of the Smithsonian
Institution.

\end{document}